\documentclass[usenatbib,usegraphicx]{mn2e}

\newcommand{\ha}{\mbox{$\mbox{\rm{H}}\alpha$}}
\newcommand{\hb}{\mbox{$\mbox{\rm{H}}\beta$}}
\newcommand{\hg}{\mbox{$\mbox{\rm{H}}\gamma$}}
\newcommand{\hd}{\mbox{$\mbox{\rm{H}}\delta$}}
\newcommand{\bg}{\mbox{$\mbox{\rm{Br}}\gamma$}}
\newcommand{\pb}{\mbox{$\mbox{\rm{Pa}}\beta$}}
\newcommand{\kmsec}{\,\mbox{$\mbox{km}\,\mbox{s}^{-1}$}}
\newcommand{\msolyear}{\mbox{$\rm{M}_{\odot}\, \rmn{yr}^{-1}$}}
\newcommand{\rstar}{\mbox{$\rm{R}_{*}$}}
\newcommand{\rsol}{\mbox{$\rm{R}_{\odot}$}}
\newcommand{\mstar}{\mbox{$\rm{M}_{*}$}}
\newcommand{\msol}{\mbox{$\rm{M}_{\odot}$}}

\author[N.H. Symington, T.J. Harries and R. Kurosawa]{Neil H. Symington, Tim J. Harries and Ryuichi Kurosawa \\
School of Physics, University of Exeter, Stocker Road, Exeter EX4 4QL}

\title{Emission-line profile modelling of structured
  T~Tauri magnetospheres}

\begin{document}

\maketitle

\begin{abstract} We present hydrogen emission line profile models of
  magnetospheric accretion onto Classical T~Tauri stars. The models
  are computed under the Sobolev approximation using the
  three-dimensional Monte Carlo radiative-transfer code {\sc torus}.
  We have calculated four illustrative models in which the accretion
  flows are confined to azimuthal curtains -- a geometry predicted by
  magneto-hydrodynamical simulations. Properties of the line profile
  variability of our models are discussed, with reference to dynamic
  spectra and cross-correlation images. We find that some gross
  characteristics of observed line profile variability are reproduced
  by our models, although in general the level of variability
  predicted is larger than that observed. We conclude that this
  excessive variability probably excludes dynamical simulations that
  predict accretion flows with low degrees of axisymmetry.

\end{abstract}


\section{Introduction}
\label{sec:intro}

Classical T~Tauri stars (CTTSs) are understood to be low-mass pre-main
sequence stars accreting from circumstellar discs. Their spectra
typically show emission lines of hydrogen and He~\textsc{i}, and both permitted
and forbidden metal lines (see e.g. \citealt{1989ARA&A..27..351B}),
while red-shifted absorption at high velocities (e.g
\citealt{1994AJ....108.1056E}) suggests infall to the stellar surface.
The basis for several models of the accretion process is a structured
stellar magnetic field disrupting viscous disc transport near the
Keplerian co-rotation radius. Gas is loaded onto the field lines and
freefalls to produce hot impact regions at mid or high latitudes. Photospheric
absorption line depths are diminished in comparison to those of
similar main-sequence stars because of the veiling by excess continuum
from the accretion hotspots. Magnetospheric accretion models are
reviewed by \citet{1998apsf.book.....H}.

The interaction between a stellar magnetic field and a disc can change the
angular momentum evolution of the star. T~Tauri stars usually rotate well below
their break-up velocity despite accreting substantial amounts of material from
a circumstellar disc.  Details of the magnetospheres predicted by different
models (e.g.  \citealt{1991ApJ...370L..39K}, \citealt{1994ApJ...429..781S})
will have different effects on the stellar rotation period. Accretion from
outside the Keplerian co-rotation radius can dissipate angular momentum to the
disc, but gas from within the co-rotation radius will spin up the star. The
requirement of a disc for this control mechanism
(\citealt{1993A&A...274..309C}, \citealt{1996MNRAS.280..458A}) could be used to
explain multiple populations of PMS stellar rotation rates (see e.g.
\citealt{1996AJ....111..283C}, \citealt{1999AJ....117.2941S},
\citealt{2002ApJ...566L..29H}). Most models assume a dipole field configuration
of order 1kG, although the finite radius over which the field must couple to
the disc will introduce some distortion due to differential rotation on either
side of the co-rotation radius. A more complex geometry with open magnetic
field lines could launch the outflows
\citep[e.g.][]{1994ApJ...429..781S,1996A&A...308...77P} that are inferred from
observations of blueshifted absorption.  \citet{1999ApJ...510L..41J} detected a
2kG longitudinal field component from a He$\,$\textsc{I} emission line in the
spectrum of BP~Tau, providing some evidence for ordered magnetic fields around
CTTSs. Further observations \citep{2003csss...12..729V,TEMPsym04b} found that
the field strength of CTTS varied smoothly over consecutive nights and might be
modulated by rotation of the system.

\citet{1994ApJ...426..669H} presented the first radiative transfer models of
CTTSs with accretion occuring via an axisymmetric dipolar magnetic field
structure, and it was found that the simulated hydrogen line profiles were
successful in reproducing the gross characteristics of the observations.
Improved models were presented by \citet{1998ApJ...492..743M} and
\citet{2001ApJ...550..944M}, incorporating line damping and extending the
statistical equilibrium calculation of the hydrogen atom to 20 levels. These
developments led to better quantitative agreement with the observed line
profiles, to the extent that the model has subsequently been used to derive
mass-accretion rates from H$\alpha$ observations (e.g.
\citealt{2003ApJ...592..266M}; \citealt{2003A&A...409..169B};
\citealt{2004MNRAS.351L..39L}).

Despite these successes, it is clear that the circumstellar geometry around a
typical CTTS is far from a static, axisymmetric, dipolar inflow. For example,
monitoring of six CTTSs over five nights (\citealt{1999MNRAS.304..367S})
revealed permitted line profile variability attributed to rotation (time-scales
of days), variable accretion (several hours), and flaring (about an hour). It
was found that the \hg\ and \hd\ line profile variability was strongly
correlated but that there was a time delay between variations in the
high-excitation lines and those from lower-energy lines. Hourly variations seen
in RU~Lup by \citet{2002A&A...391..595S} are attributed to varying mass
accretion rates and/or changes in the distribution of accreting matter. Those
authors conclude that the observations rule out an axisymmetric accretion flow
from a disc aligned perpendicular to the stellar rotation axis.  A
spectroscopic and photometric study of AA~Tau (\citealt{2003A&A...409..169B})
revealed a time delay between variations in the \ha\ and \hb\ profiles and the
veiling continuum. It was argued that this lag was the result of a change in
accretion rate propagating (in freefall) along the field lines, with the change
in density being observed first in the line profiles and then in the continuum
when the increased kinetic energy of the flow is liberated at the hot spots.

Variability is often observed on rotational time-scales.
\citet{1995ApJ...449..341J} obtained optical time-series spectra of SU~Aur and
found variations in the red-shifted absorption component of \hb\ on the same
time-scale as the stellar rotation period. A blue-shifted absorption feature
consistent with an outflow was found to be anti-correlated with the red-shifted
inflow signature, suggesting a magnetosphere that is seen to alternately favour
inflow then outflow at different rotational phases.
\citet{1996A&A...314..821P} also found evidence for simultaneous outflow and
infall in SU~Aur (Balmer lines and He~\textsc{i} 5876\AA).  The possibility of
an offset dipolar accretion structure was considered, but the complex
kinematics of the magnetically channelled gas and the degeneracy of the
observational diagnostics make quantitative conclusions difficult to achieve.

Long-term monitoring of CTTS TW~Hya (\citealt{2002ApJ...571..378A}) revealed
evidence for both inflow and outflow in the Balmer line profiles, with the
deepest \ha\ blueward absorption occuring at phases when the veiling continuum
(and hence mass accretion rate) is greatest. It was suggested that the system
is viewed at low inclination, although the presence of photometric variability
argued against a truly face-on orientation. A similar viewing angle was used to
explain the redwards peaked Balmer line profiles of DR Tau
(\citealt{2001AJ....122.3335A}).

\citet{2000A&A...362..615O} performed cross-correlation analysis of multiple
spectral lines from SU~Aur to investigate non-coincident changes.  Different
components of line profiles were seen to be time-lagged with respect to each
other, which is inconsistent with the expectation for a simple geometry. It was
suggested that the accretion flow had an azimuthal component and that the
observer's line of sight was sampling different different physical environments
for line formation.  Structured accretion is also supported by the observations
  of VZ~Cha \citep{2001A&A...378.1003S} who report features that persist for
  longer than the likely freefall time from the inner edge of a truncated disc.
  The possibility of confinement to two rotating streams was considered, with
  the observed period being only half the stellar rotation period.

If we are to fully understand the magnetospheric accretion process,
from its role in rotational breaking to the use of line profiles as
accretion diagnostics, we must obtain a better description of the
geometry of the circumstellar flows. The observational evidence for
structured accretion is overwhelming, and it is now appropriate to
challenge these data with theoretical models. In this paper we present
the first 3D radiative-transfer modelling of azimuthally-structured
accretion. We note that it is not the aim of this work to reproduce in
detail the variability of permitted line profiles in CTTS, but rather
to gain insight into the likely characteristics of line profile
variability that would be produced by `cartoon' geometries that have
been invoked by other researchers to interpret their timeseries data.
As a first step we adopt a canonical, bipolar reference model, and
examine the synthetic line profiles produced by simple departures from
axisymmetry in the magnetosphere.

\section{Radiative transfer modelling}
\label{sec:radiative}

We have extended the \textsc{torus} radiative-transfer code
\citep{2000MNRAS.315..722H} in order to simulate hydrogen line
profiles from pre-main-sequence stars with magnetospheric accretion
flows. The circumstellar environment of accreting T~Tauri stars
encompasses a large range of length scales. Assuming that the material
is strongly influenced by the magnetosphere, we must consider a volume
extending to several stellar radii from the surface.  The addition of
a disc (or a large-scale outflow) will further enlarge the space in
which our simulated photons may have significant interactions. Close
to the stellar surface, the magnetic fields may confine gas in narrow
columns, leading to small impact regions at the photosphere. The
liberation of energy by shock heating by the infalling gas in these
areas may be the source of much of the system's luminosity.

The complexity of the circumstellar environment requires a careful
choice of mapping to the simulation's grid. Regular subdivision of our
model space (e.g. in spherical polar coordinates) leads to many of the
cells representing regions with little effect on the radiation field.
By adopting the technique of adaptive mesh refinement (AMR) we can
tailor the subdivision of the simulation space to improve the sampling
of the structures it contains \citep[see
e.g.][]{2001A&A...379..336K,2003A&A...401..405S, 2004MNRAS.350..565H,
  2004MNRAS.351.1134K}.  Our whole cube-shaped space is divided into
eight, smaller cubic subcells; and those subcells are themselves
repeatedly divided until some criterion is satisifed throughout the
grid. For the models presented in this paper, we specify the maximum
mass that any subcell can contain. This rule can be applied
efficiently if we can quickly calculate the density at any location
within the grid; and it concentrates many small subcells where gas is
present, while the empty regions are only sampled by a few large
cells. \citet{NHSthesis} gives further details of our AMR
implementation.

\begin{figure}
\centering
\begin{tabular}{c}
\includegraphics[width=8cm,angle=0,clip=true]{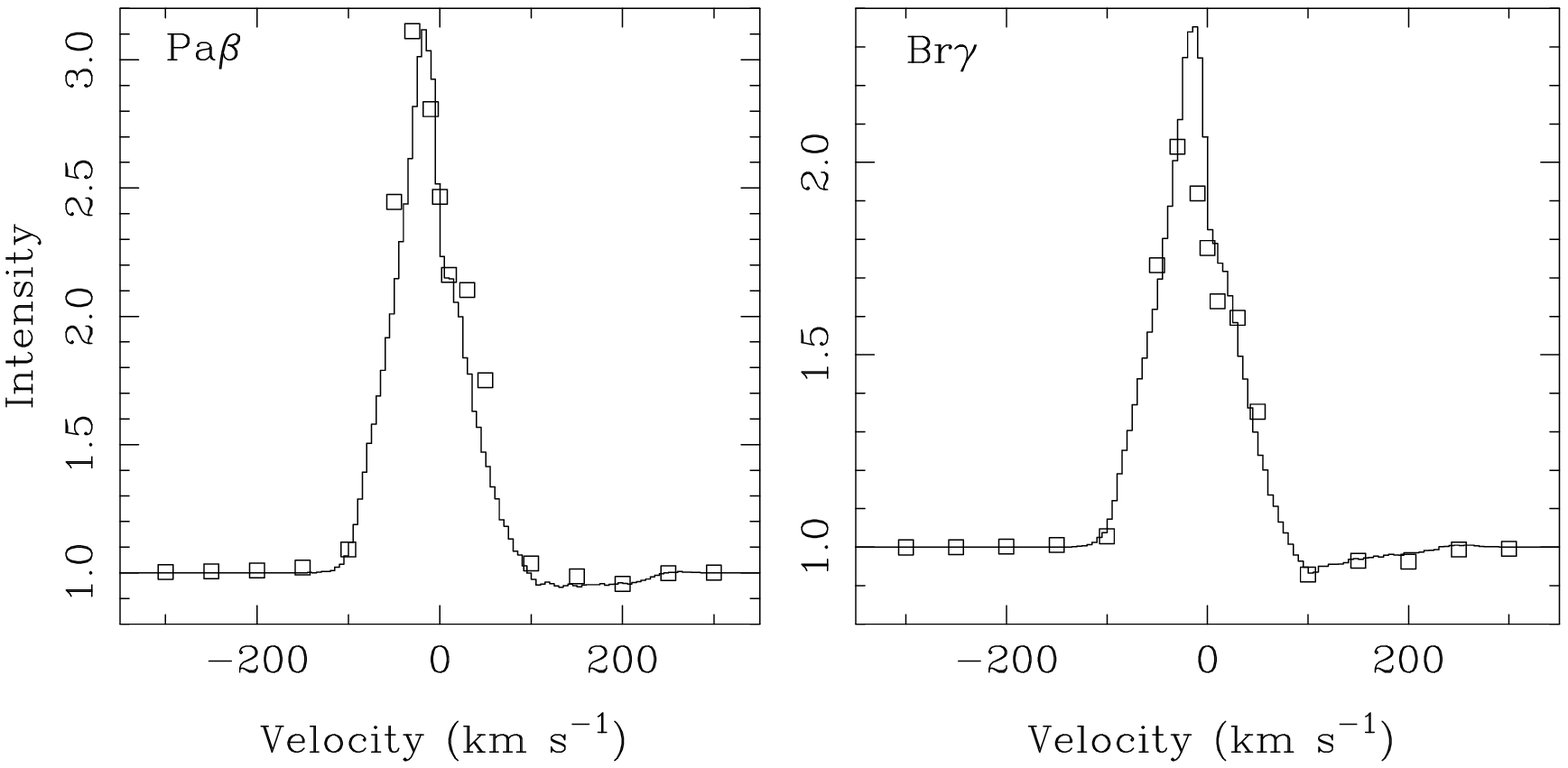}
\end{tabular}

\caption{Hydrogen line profiles for an axisymmetric accretion geometry
  with parameters: $\dot{{\rm M}}= 10^{-7} \msolyear$, accretion stream
  $T_{\rm max}=7000\rm{K}$, $\rstar=2\rsol$, $\mstar = 0.5\msol$,
  viewing inclination = $30\degr$. The accretion stream begins in the
  disc plane between $\rm{r}_{\rm inner}=2.2\rstar$ and
  $\rm{r}_{\rm outer}=3.0\rstar$. Solid line: \textsc{torus} 3-D computation
  with AMR scheme. Squares: MCH computation (reproduced from
  \texttt{http://cfa-www.harvard.edu/cfa/youngstars/}) }
\protect\label{fig:compare}

\end{figure}

\subsection{Accretion flow model}
\label{sec:flowmodel}

In order to investigate the effect of azimuthal structure on the line
profiles we must adopt a reference model for the circumstellar
geometry. Here we use the model of an accretion disc and stellar
magnetic field given by \citeauthor{1994ApJ...426..669H} (\citeyear{1994ApJ...426..669H}, hereafter HHC),
in which a dipole magnetic field intersects the disc and allows gas to
free-fall along the field lines to the stellar surface. The pressure
of the gas picked up from the disc is assumed to cause no distortion
of the magnetic field, and the innermost field line defines the radius
where the disc is completely truncated. The density and velocity at
any point in the flow can by computed directly from the analytical
formulae of HHC, which are based on the conservation of mass and
energy.  We have not attempted to include rotation of the
magnetosphere in our model.  This simplification may slightly change
the central peak of our simulated line profiles (dominated by low
velocity gas near the disc), but should not affect the line wings (see
\citealt{2001ApJ...550..944M}, hereafter MCH). Although not as
sophisticated as using a Shu-type magnetosphere
(\citealt{1994ApJ...429..781S}), or indeed adopting the density
structure from an MHD simulation (e.g. \citealt{2003ApJ...595.1009R}),
this is the simplest, plausible circumstellar structure, and it
enables us to easily investigate the effect of departures from
axisymmetry on the emission lines.

Upon reaching the stellar surface, shock heating of the infalling gas
will liberate its kinetic energy. We adopt the assumption of HHC that
the radiation (typically X-rays) is immediately reprocessed and all
the accretion energy is radiated from the impact region as a
blackbody. The geometry and mass-transfer rate of our accretion flow
(both of which quantities may vary with time) define the pattern of
hot-spots over the stellar surface. At each temporal phase in our
simulations we calculate the kinetic energy of the flow directly above
a grid of points distributed over the stellar surface. Ignoring any
heating/cooling time-scale, we specify the emitted spectrum of each
surface element to be the sum of the intrinsic stellar spectrum and a
blackbody contribution from the instantaneous local accretion rate.
The time-independent accretion models presented in this paper create
hot-spots with temperature $\sim7250\mathrm{K}$.

Perhaps the greatest uncertainty in the magnetospheric
radiative-transfer modelling to date is the temperature structure of
the accretion flow. The form of the temperature run will have a
significant impact on the line source functions, and hence the line
profiles themselves. To our knowledge the only study to address this
aspect explicitly is the work of \citet{1996ApJ...470..537M}. He
calculated the temperature structure of dipolar accretion flows with
the same geometry and density as HHC. Heating was found to be
dominated by adiabatic compression as the magnetic field lines
converge along the path of the flow, while the major coolants were
bremsstrahlung radiation and line emission from Ca\,{\sc ii} and
Mg\,{\sc ii}. He found that the temperature of the accretion flow grew
from the disc, and reached peak temperatures of $\sim 6000$\,K before
the temperature fell as the density (and hence the cooling) increases
towards the stellar surface.  For the purposes of this paper, in which
we are concentrating geometrical effects, we have adopted the
temperature profile of HHC (see fig. 6 in their paper), which
qualitatively matches the \citet{1996ApJ...470..537M} results. We
defer studying departures from this temperature law to future work,
but note that multi-line spectral analysis of observations should
provide strong constraints on the magnetospheric temperatures.

The statistical equilibrium code is based on \textsc{stateq}
\citep{1995PhDT.......168H}, an implementation of the method of
\citet{1978ApJ...220..902K}. The radiation field is computed by
integrating the spectrum from each surface element of the star,
weighted by the solid angle subtended from the centre of a grid cell.
We typically take advantage of any symmetries present in our models to
increase the computational efficiency. For example, for axisymmetric
accretion flows, we can initially calculate the statistical
equilibrium for grid cells in a plane, and then rotationally map the
solution characteristics to the other cells. The iterative
calculations for all of the cells throughout the 3D grid will then
proceed quickly because they are starting close to the solution they
will converge upon. We compute the first 14 levels of the hydrogen
atom in order to have sufficient high-$n$ levels to avoid the
inevitable overpopulation of levels near the continuum impacting on
the upper levels of the lines under consideration.

\begin{figure}

  \center \begin{tabular}{l}
  \includegraphics[width=8cm,clip=true]{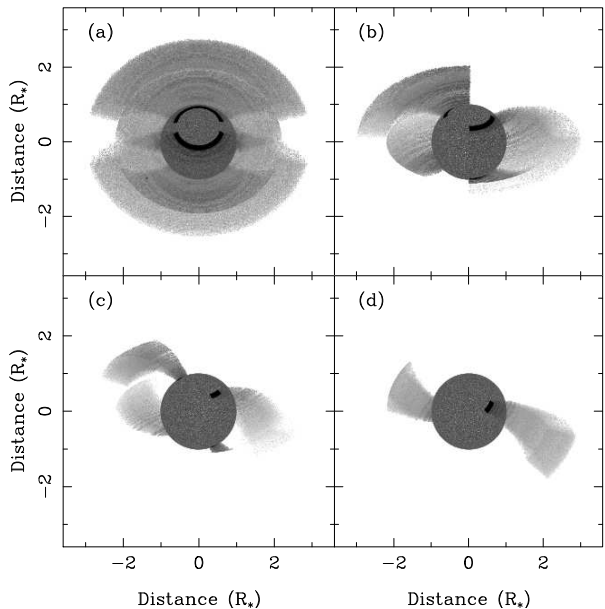}
  \end{tabular} 

  \caption{\label{fig:images1}   The four accretion models (see
  Table~\ref{table:models}): (a) D150  (b) D90 (c) D30 (d) A30. The observer
  views the systems at inclinations of 30\degr\,(a and d) or 60\degr\,(b and
  c). The intensity is measured over a narrow  wavelength band around the
  \ha\ line (6550--6575\AA). The accretion footprints are obvious  as bright
  hotspots at high latitudes. The images are logarithmically-scaled over three
  decades from most intense (black) to least intense (white). The opaque
  accretion discs truncate the  view of the accretion stream(s) below the disc
  planes. }
\end{figure}

\subsection{Monte Carlo radiative transfer}
\label{sec:montecarlo}

The radiative transfer algorithm is described in
\citet{2000MNRAS.315..722H} with reference to a spherical polar grid.
We have updated the implementation to use recursive routines so that
it can be applied to our AMR grid, which is stored in a tree
structure. The emissivity of each grid cell in the flow region is used
to weight the probability of that cell being the source of a Monte
Carlo photon. The stellar photosphere is separately represented by a
grid of small surface elements. Photons originating from the star are
weighted by the continuum luminosity of each element at the wavelength
of the spectral line.

The numerical integration from the source of a photon to the observer
is described in \citet{2000MNRAS.315..722H}. Sampling of the
conditions along each photon's path is done with a recursive traversal
of the tree structure storing the AMR cells. The line transfer is
performed under the Sobolev approximation, and we neglect non-local
coupling terms (which would render our 3D calculations computationally
intractable). Both assumptions lead to a negligible effect in the line
wings, but may have a significant impact at the line core.
Additionally, we do not include the effects of line damping. Our model
profiles are therefore likely to underestimate the true extent of the
line breadth for \ha\ (see MCH for a discussion and comparison of
results) but will have little effect on the higher order lines. 

Alfen-driven turbulence has been invoked to explain the broad,
symmetric wings that are observed in some profiles (e.g.
\citealt{1981ApJ...244..124D}; \citealt{1982ApJ...261..279H};
\citealt{1990MmSAI..61..707B}; \citealt{2000AJ....119.1881A}). We have
not included turbulent broadening in the calculations presented here,
since if the velocity amplitude of the turbulence is comparable with
the bulk flow velocity then the Sobolev approximation breaks down. As we
are interested in the imprint of geometrical structure on the emission
line profiles we do not regard the lack of turbulent broadening in our
models as a significant problem, and furthermore we note that its
inclusion requires co-moving statistical equilibrium calculations in three-dimensions, a
problem that is computational intractable at present.

We choose to present our synthetic spectra normalized to the local
continuum at each time interval. We do not correct for veiling of
absorption features or suppression of emission features by a changing
continuum level. While observations of comparable spectra can
compensate for these effects \citep[e.g.][]{1994AJ....108.1056E},
these corrections introduce their own uncertainties and many authors
present spectra without such processing
\citep{2001AJ....122.3335A,2001A&A...378.1003S}.

\subsection{Code testing}
\label{sec:codetesting}

We have performed numerical tests to ensure the spatial resolution of
our simulations (typically based on $\sim 10^{6}$ grid cells) is
sufficient to represent the accretion flow geometry we are using.  In
Fig.~\ref{fig:compare} we compare \pb\ and \bg\ profiles (selected to
minimize discrepancies due to line damping) computed in 3D using {\sc
  torus} with the profiles for the same geometry calculated using the
code described by MCH. Given the slightly different physics underlying
the two codes the profiles show satisfactory agreement in intensity
and morphology.

The Monte Carlo radiative transfer data sets presented later have each
been generated from at least $1\times 10^6$ photon packets. The noise
correlation between our spectral wavelength bins, and our adoption of
many variance-reduction techniques, prohibits simple estimation of the
effective signal-to-noise ratio, but the excellent reproducibility of
our results suggests noise is insignificant.

\begin{figure*}
  \center \begin{tabular}{l}
  \includegraphics[width=14cm,angle=0,clip=true]{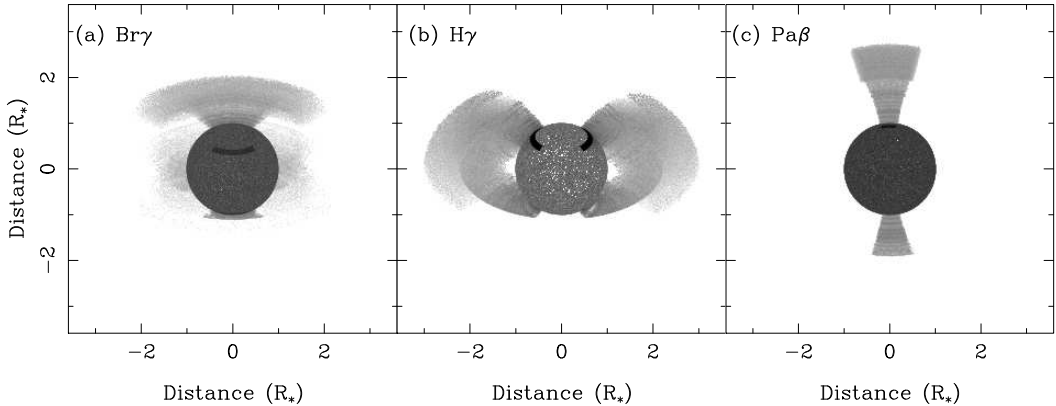}
  
  \end{tabular} \caption{\label{fig:images2} Intensity images of accretion
  models at selected phases. Each image shows flux integrated over a narrow
  wavelength range (corresponding to $\pm500\kmsec$) around the rest wavelength
  of the indicated spectral line. (a) D90 viewed at the wavelength of \bg\ at
  phase 0 ($60\degr$\,inclination). The two streams lie directly in front and
  behind the star. (b) A view of D90 at phase 0.25 (\hg\ passband,
  60\degr\,inclination). The streams are perpendicular to the line of sight.
  (c) A30 at \pb\ at phase 0.5 (30\degr\,inclination). The lower stream is
  towards the observer but viewed through the disc's inner hole and so appears
  truncated. The upper stream is on the far side of the star -- the accretion
  hotspot is marginally visible at the limb. The images are logarithmically
  scaled over four decades, from black (most intense) to white (least
  intense).}

\end{figure*}

\subsection{Accretion models}
\label{sec:example}

Time-series observations of the line profile variability of CTTSs have not yet
provided unambiguous clues to the structure of the accretion flows. The
observer is likely seeing a geometry with overall characteristics persisting
for longer than the rotational time-scale, but with some structure varying more
  rapidly. The accretion rate is also expected to show changes on short
  time-scales dependent on the changing magnetic field configuration
  \citep{2001A&A...378.1003S}.  

In this paper, we aim to identify some of the characteristics of the
magnetically-controlled accretion. If, in one simulation, we depart in too many
ways from the simple steady-state models it will be difficult to identify the
contributions from each change. We have therefore made only small adaptations
to the HHC magnetospheric accretion geometry, breaking the axisymmetry to
investigate line profiles as the star rotates. The accretion flows are confined
to two azimuthal curtains around the star so that their projection against the
stellar disc and hot-spots changes during a rotation cycle. This scenerio was suggested by
\citet{2001A&A...378.1003S} as an explanation for the periodicity in the
spectra of VZ~Cha. 

This configuration also is similar to that found in the 3D MHD
simulations of \citet{2003ApJ...595.1009R}. Those authors model a
magnetic field slightly inclined relative to the star's rotation axis
and the disc plane. Accretion is energetically favourable at some
azimuth positions, and the flow becomes concentrated in two streams
(see fig.  5 of their paper). We have simulated accretion curtains
which are symmetric about the disc plane, and also streams that are
diametrically opposite about the star.

\begin{table*}
  \caption{\label{table:models}
  Model parameters and properties of the spectral continua. Double accretion
  curtains are mirrored about the disc plane, while the alternate configuration
  has one stream above the disc, and one below, at opposing azimuth angles.
  These models are illustrated in Fig.~\ref{fig:images1}. The mean levels are
  quoted for an observer at a distance of $100~\mathrm{pc}$. The range is the
  full range of the continuum variability, expressed in magnitudes.  The final
  column gives the relative contribution to the continuum flux from the
  accretion hotspot ($\mathrm{f}_{\mathrm{acc}}$), relative to the photospheric
  flux ($\mathrm{f}_{\mathrm{phot}}$) at the line wavelength.} 
  
  \center

  \begin{tabular}{lccccccccc}
  \hline

  Model & \multicolumn{2}{c}{Accretion curtains} & $\dot{\rmn{M}}$ & Magnetic offset &
    Line & Inc. & Mean continuum   & Range & $\mathrm{f}_{\mathrm{acc}}/\mathrm{f}_{\mathrm{phot}}$\\   
    
     & Type & Angular size & (\msolyear) &  &  & &($\mathrm{erg}\,\mathrm{s}^{-1}\,\mathrm{cm}^{-2}\,\mathrm{\AA}^{-1}$) & (mag) &  \\ 
  \hline

  D150 & Double & 150\degr  & $8.3\times 10^{-8}$  & 0\degr & \hg & 60$\degr$ &$3.9\times 10^{-13}$ & 0.06 & 3.9 \\ 

  D90  & Double & 90\degr   & $5.0\times 10^{-8}$  & 0\degr & \hg & 30$\degr$ & $3.7\times 10^{-13}$ & 0.0 & 2.3  \\ 
       &        &           &                      &        &     & 60$\degr$ & $2.7\times 10^{-13}$ & 0.15 &      \\ 
       &        &           &                      &        & \ha  & 60$\degr$ & $3.4\times 10^{-13}$ & 0.07& 0.58 \\
       &        &           &                      &        & \pb  & 60$\degr$ & $1.7\times 10^{-13}$ & 0.03 & 0.16  \\
       &        &           &                      &        & \bg  & 60$\degr$ & $5.7\times 10^{-14}$ & 0.002& 0.08   \\
        
  D30  & Double & 30\degr   & $1.7\times 10^{-8}$  & 0\degr & \hg  & 60$\degr$ & $1.5\times 10^{-13}$ & 0.14& 0.78  \\ 
  
  A30 &Alternate& 30\degr   & $8.3\times 10^{-9}$  &10\degr & \hg  & 30$\degr$ & $1.3\times 10^{-13}$ & 0.3& 0.38  \\ 
      &         &           &                      &        & \ha  & 30$\degr$ & $2.5\times 10^{-13}$ & 0.13& 0.096 \\
      &         &           &                      &        & \pb  & 30$\degr$ & $1.5\times 10^{-13}$ & 0.04& 0.026 \\
      &         &           &                      &        & \bg  & 30$\degr$ & $5.3\times 10^{-14}$ & 0.03& 0.013 \\

  \hline
  \end{tabular}
\end{table*}

\citet{2003ApJ...595.1009R} found that the accretion was axisymmetric in the
case where the magnetic and rotation axes were the same. For one of our
non-axisymmetric models, we have therefore implemented an offset between these
two axes. The HHC formulation for the accretion flow would not represent such a
misalignment without substantial reworking. We have instead used the HHC
accretion flow unmodified, and achieved the offset effect by only changing the
rotation axis for our simulated observations, and the disc axis. The start of
the accretion flow is then not coincident with the surface of the disc, but the
results are approximately valid for small angular offsets (we choose 10\degr). 

We adopt the CTTS system parameters of HHC: $\mstar = 0.8\msol$,
$\rstar=2\rsol$; and an accretion stream beginning close to the disc plane
between $\rm{r}_{\rm inner}=2.2\rstar$ and $\rm{r}_{\rm outer}=3.0\rstar$. The
stellar continuum flux is described by a model atmosphere for $\mathrm{T}_{\rm
eff}=4000\mathrm{K}$, $\mathrm{log\:g}=3.5$ \citep{1993KurCD..13.....K}.  The
mass accretion rate \textit{per unit azimuth angle, per hemisphere} is the same
in each case (integrating to $10^{-7}\rmn{M}_{\odot}\,\rmn{yr}^{-1}$ if the
accretion streams were complete), but the limited extent of the flow in each
model reduces the effective accretion rate. Table~\ref{table:models} shows the
parameters that differ between the models.  We have simulated line profiles at
viewing inclinations measured with respect to the stellar \textit{rotation}
axis. 

\section{Results}
\label{sec:results}

The results we present are grouped to demonstrate the effects of changing
individual aspects of the simulation (spectral line, accretion geometry and
viewing inclination). For two of the models, we show time-series spectra
simulated for four spectral lines (Sections~\ref{sec:D90} and \ref{sec:A30}).
We also compare the spectra for a single line (\hg) as the accretion geometry
or viewing inclination are altered (Sections~\ref{sec:extents} and
\ref{sec:incs} respectively).

\begin{figure*}
\begin{minipage}{180mm}
  \centering
  \begin{tabular}{c}
  \includegraphics[width=18cm,angle=270,clip=true]{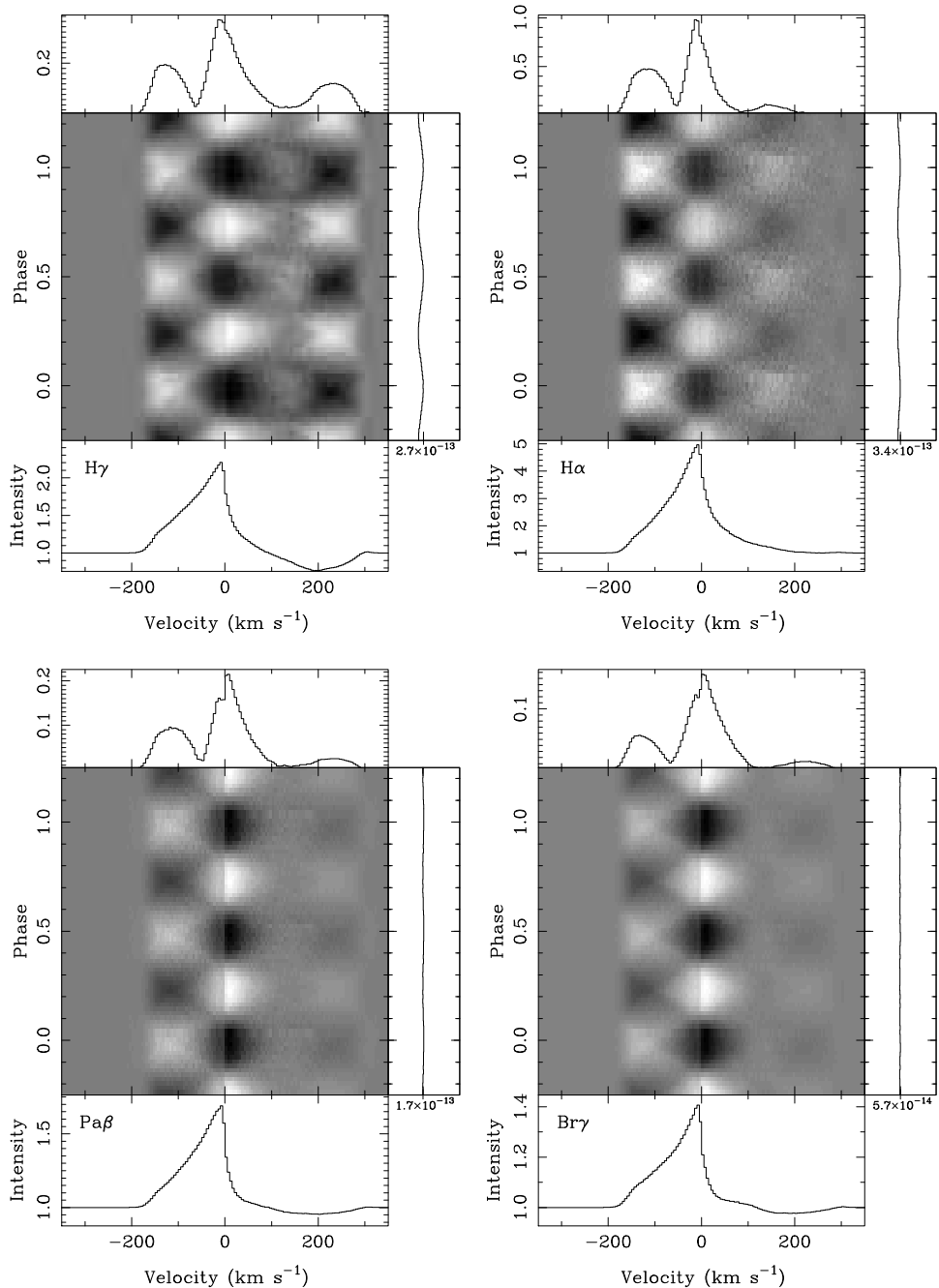}
  \end{tabular}
  
  \caption{Spectra from D90 model, viewed at 60\degr\ inclination. The bottom
  plots show the mean spectrum for each hydrogen line. The greyscale image
  shows the quotient of the spectrum at each rotational phase, with respect to
  the mean spectrum.  The variability shown in the greyscale has ranges (from
  top left) of 57, 79, 47 and 35 per cent of the mean spectrum in a given
  wavelength bin. The plot beside each greyscale image shows the continuum level
  at each phase -- the scales are presented in Table~\ref{table:models} and
  the values increase towards the right hand side. The RMS deviations of each
  wavelength bin are plotted in the uppermost box, with the zero point at the
  bottom of the box.}
  
  \protect\label{fig:d90}
\end{minipage}
\end{figure*}  

\subsection{Spectral line comparison}
\label{sec:D90}

The D90 model has accretion streams on opposite sides of the star, both above
and below the disc, i.e. four in total (Figs.~\ref{fig:images1}b, 3a, 3b).  Each stream
covers an azimuth range of 90\degr, and the whole system is viewed from an
inclination of 60\degr. At this angle, the upper accretion hotspots alternately
rotate out of the line of sight, modulating the oberved continuum intensity
(Fig.~\ref{fig:d90}; Table~\ref{table:models}). This effect is most pronounced
(0.15 mag) at the wavelength of the \hg\ line  because the hotspots contribute
most significantly to the blue continuum. 

\begin{figure}
\centering
\begin{tabular}{c}
\includegraphics[width=8.0cm,angle=0,clip=true]{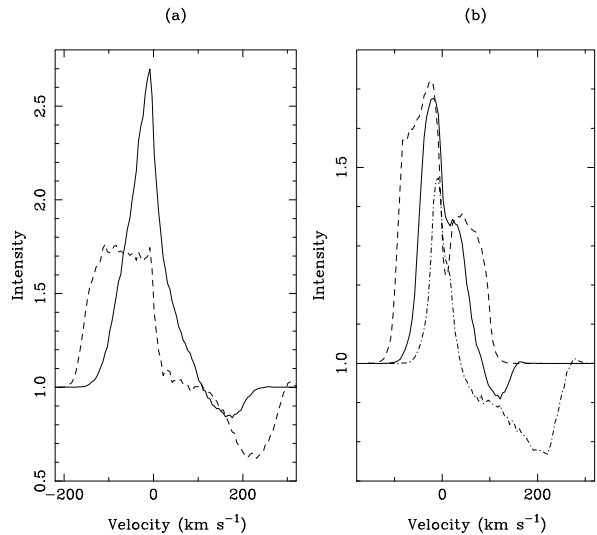}
\end{tabular}

\caption{\hg\ spectra at selected phases. (a) D90 model (viewed at $i=60\degr$)
when curtains perpendicular to line of sight (phase 0.25; solid line) and along
line of sight (phase 0; dashed line). (b) A30 model (viewed at $i=60\degr$)
with curtains perpendicular to line of sight (phase 0.25 or 0.75; solid line),
with upper stream on far side of star (phase 0.5; dashed line) and with upper
stream towards observer (phase 0; dot-dashed line).}
\protect\label{fig:snap_plots} \end{figure}

\begin{figure}
  \centering
  \begin{tabular}{c}
  \includegraphics[width=8cm,clip=true]{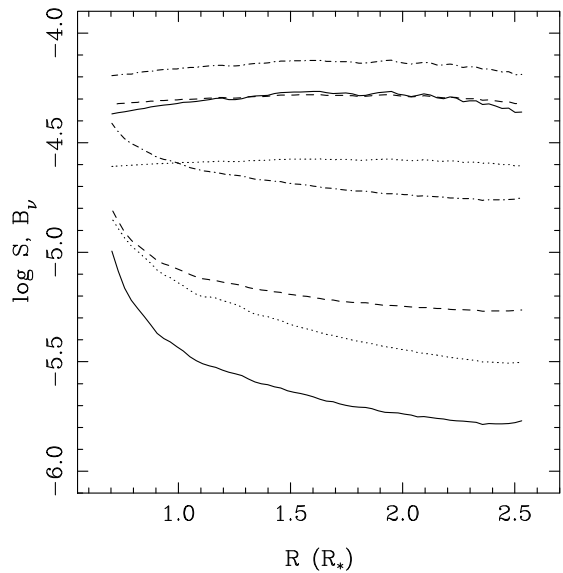}
  \end{tabular}
  
  \caption{Source functions, S, for spectral lines in the accretion columns.
  Solid line: \hg; dot-dashed: \ha; dashed: \pb; dotted: \bg. The local Planck
  function $B_\nu$\ is shown in the same style. For a given spectral line, the
  upper line shows $B_\nu$\ and the lower line shows S.}
  
  \protect\label{fig:source}
\end{figure}

The hydrogen line profiles show periodic variations as the projected velocity
of the inflowing streams changes over a rotational period (Figs.~\ref{fig:d90},
\ref{fig:snap_plots}a). The mean spectral profiles are all peaked at the rest
wavelength of the line and are asymmetric with most of the emission occuring
blueward of line centre. \ha\ is the strongest of the lines and is in emission
over its full width; the others all show inverse P~Cygni profiles, with \hg\
having the most significant red-shifted absorption feature. The origins of the
line profile shapes are found in the line source functions
(Fig.~\ref{fig:source}). The source functions of all four lines lie below their
local Planck functions, although the lines become more thermalized as the
density increases rapidly close to the stellar surface. For \hg\ and the near
infrared transitions, the line source functions are significantly smaller than the Planck function for
a $T\sim 7250$K continuum source: an IPC profile therefore results when the hot spots are viewed
through the intervening accretion streams. The H$\alpha$ source function is significantly more thermalized, 
and does not show red-shifted absorption.

Observers often present the temporal variance spectrum (TVS) of their line
profile timeseries (e.g. \citealt{1995AJ....109.2800J,2000A&A...362..615O}),
and it is a useful diagnostic for quantifying both the level and velocity
distribution of line profile variability. In Fig.~\ref{fig:d90} we present the
root-mean-square (RMS) spectra of our models, which is a comparable diagnostic
to the TVS. It is immediately apparent that the H$\alpha$ model shows the
strongest variability, and that all lines show a multiply-peaked RMS spectrum
with the highest variability occurring at line centre.

The \hg\ profile varies at three velocity ranges, two of which are well
correlated with each other (seen in the RMS plot as bands centred at $200$ and
$0\kmsec$); and the other is anticorrelated with those two (seen at
$\sim-130\kmsec$). There are two regions between these bands which are almost
invariant because the integrated line emission seen at some velocities remains
constant throughout the rotation period. \ha\ shows the same zero-velocity and
blueshifted bands, but the high-velocity red-shifted band is constant (\ha\
does not show an IPC absorption feature). In the two infrared lines, the
red-shifted and blue-shifted emission is less variable than for \hg, but the
line centre still shows large changes in intensity. The maximum of the
low-velocity component occurs at phases 0.25 and 0.75 when the streams are
perpendicular to the line of sight, and the projected velocity of the inflowing
gas is minimized (Fig.~\ref{fig:images2}b).  At these phases, all the line
profiles (Figs.~\ref{fig:d90},~\ref{fig:snap_plots}a) are composed of a
symmetric emission peak around $0\kmsec\ $, and (excepting \ha) a small
red-shifted absorption feature (only a very small volume of gas is seen in
front of the hot-spot with this configuration). 

Phases 0 and 0.5 (Figs.~\ref{fig:images2}a,~\ref{fig:snap_plots}a) have the
streams aligned with the observer, and the projected velocities of the gas are
maximized. Considering first \hg, the stream in front of the star is
responsible for strong absorption at $\sim+220\kmsec$. Blueward of line centre
is a broad flat-topped emission feature extending to $\sim-170\kmsec$. The line
profile variability in the red line wing is strongest for \hg, since the hot
spot emission (which is dominant at short wavelengths) is absorbed by the
infalling material. The red-wing variability is considerably smaller for the
other transitions, mainly due to the smaller contribution of the hot-spots to
the total continuum flux.

\subsection{Geometry comparison}
\label{sec:extents}

\begin{figure*} \begin{minipage}{180mm} \centering \begin{tabular}{c}
\includegraphics[width=18cm,angle=0,clip=true]{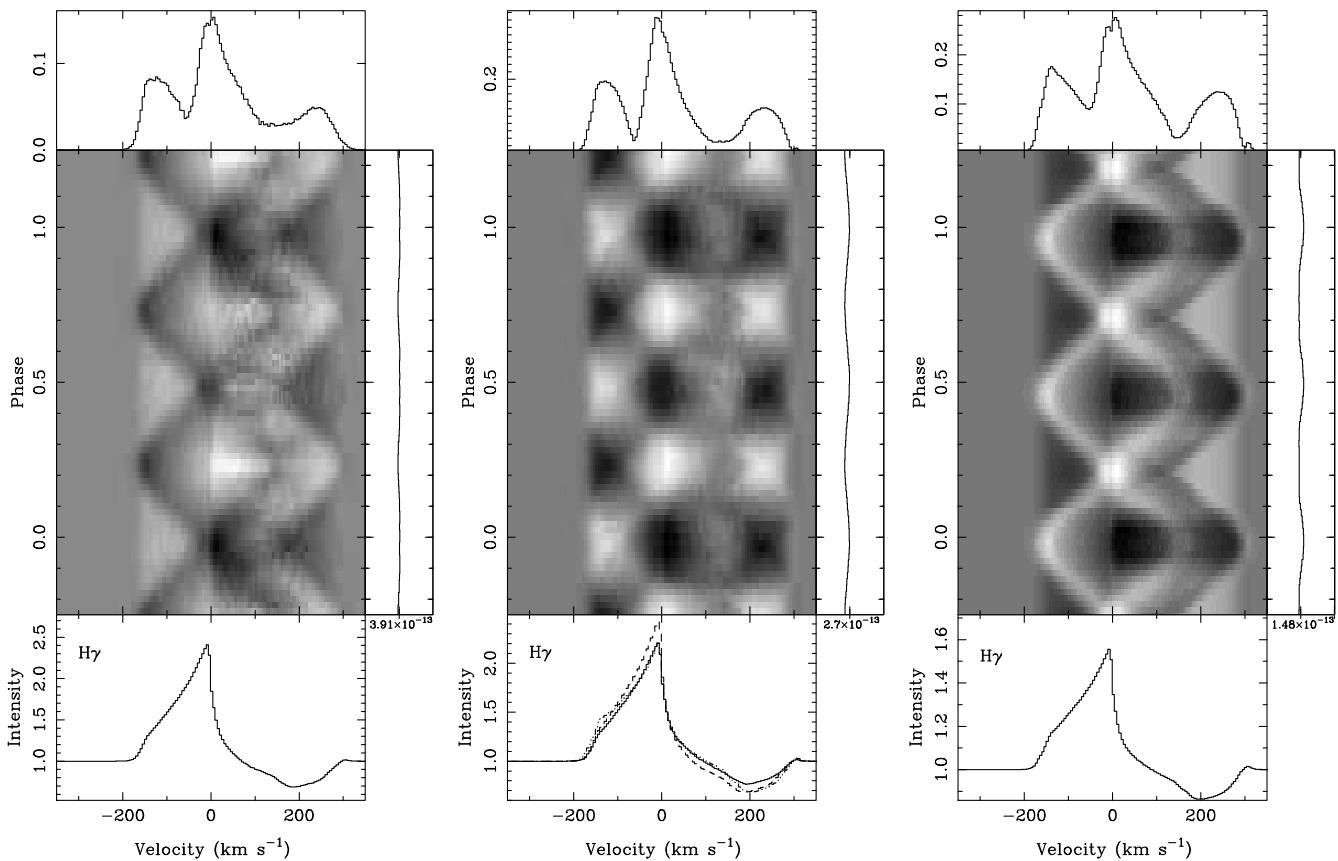}

\end{tabular} \caption{H$\gamma$ spectra for three models viewed at 60$\degr$
inclination, see Fig. 4 for a description. From left to right, D150, D90, D30. The variability shown in the
greyscale has ranges of (from left to right) 39, 57 and 65 per cent of the mean
spectrum in a given wavelength bin. The plot of the D90 mean spectrum (solid
line) also shows the D150 (dashed line) and D30 (dotted line) mean spectra
after they were all rescaled to have the same equivalent width.  } 

\protect\label{fig:allExtents} \end{minipage} \end{figure*}

Our second group of results (Fig.~\ref{fig:allExtents}) are all for the \hg\
line observed from an inclination of 60\degr, and we show synthetic spectra for
accretion streams with different azimuthal extents. The mean spectrum for the
D90 model is shown with the other two mean spectra (after rescaling to equal
equivalent widths) for comparison, and the lines have almost the same profile.
The range of variability in the continuum is directly related to the extent of
the accretion footprints, with the models with the most localized accretion
showing the most significant continuum modulation.

Although the mean spectrum can not be used to differentiate between different
models, the time-series quotient spectra could be used as an observational
diagnostic. The periodic behaviour of the lines is approximately similar in the
three simulations, but the differing filling factors of the accretion flow
produce different variations. When the narrowest accretion flow, D30
(Fig.~\ref{fig:images1}c)), is seen perpendicular to the line of sight (phase
0.25 or 0.75) there is very little gas projected against the hot-spots, and the
line has only a small absorption at $\sim120$\kmsec. At the same phases, the
D150 model (Fig.~\ref{fig:images1}a) creates an absorption feature at
$\sim180$\kmsec\ because the flow extends further in azimuth towards the
observer and has components with higher projected velocity. The emission peak
of the line profile shows equivalent behaviour for the same reason. 

With only the accretion filling factor differing between the three models, the
system's rotation will average out the contribution from high and low projected
gas velocities seen at different phases. It is therefore not surprising that
the mean spectra differ only by a scaling factor. It is clear that observations
of both line and continuum variability (that adequately sample the stellar
rotation period) are needed to distinguish between the different geometries we
have simulated here.

\begin{figure*}
\begin{minipage}{180mm}
  \centering
  \begin{tabular}{c}
  \includegraphics[width=13cm,angle=270,clip=true]{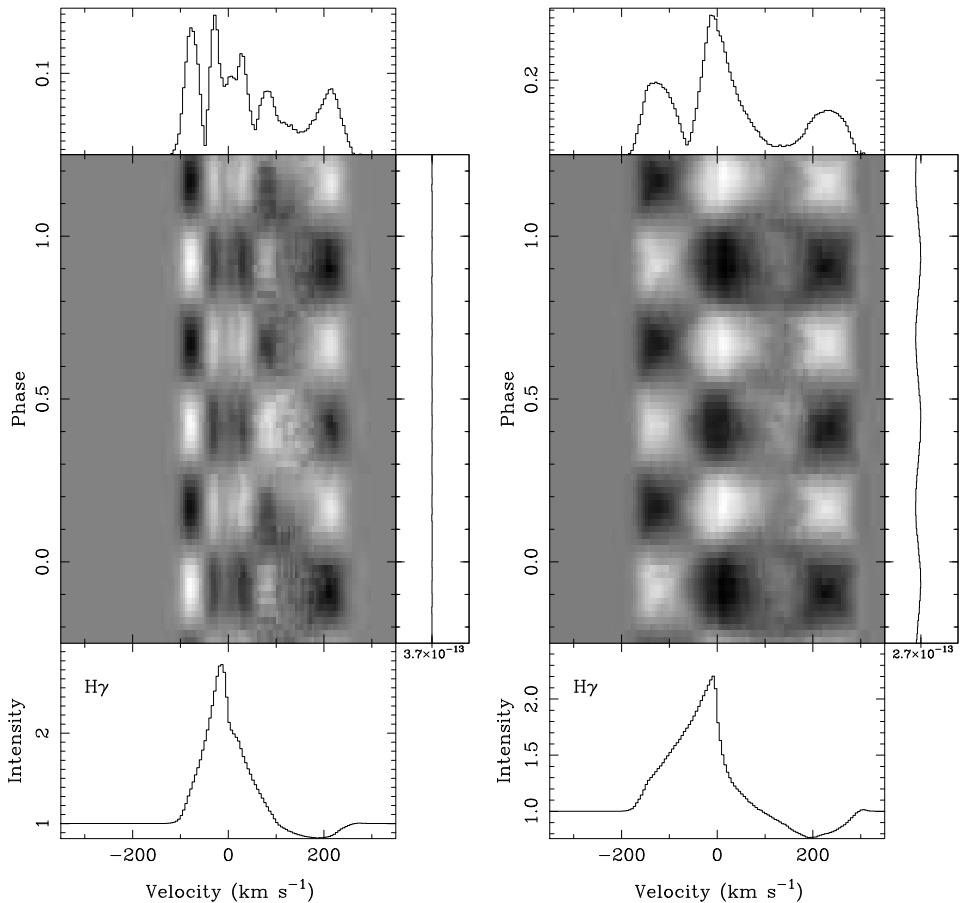}
  \end{tabular}
  
  \caption{H$\gamma$ spectra of the D90 model viewed at inclinations of
  30$\degr$ (left) and 60$\degr$ (right). See Fig.~\ref{fig:d90} for a
  description. The variability shown in the greyscale has ranges of 35 (left)
  and 57 (right) per cent of the mean spectrum in a given wavelength bin.}
  
  \protect\label{fig:incs}
\end{minipage}
\end{figure*}

\subsection{Inclination comparison}
\label{sec:incs}

Fig.~\ref{fig:incs} shows the D90 model viewed at both 30\degr\ and 60\degr\
inclination. On average, the accretion streams have larger velocity components
parallel to the disc plane than perpendicular to it, so the line profile
extends to higher velocities ($-180$--$300$\kmsec) at 60\degr, compared to
30\degr\, inclination ($-120-250$\kmsec). At 30\degr\ inclination, the
continuum variation is comparable with the photon noise, but at 60\degr one of
the upper accretion hot spots will not be visible at some phases and so the
variability becomes significant (0.15 mag).  Additionally, the two lower
hotspots are never visible from 30\degr, but at 60\degr\ one of them will be
seen at some phases. 

The overall characteristics of the time-series quotient spectra are the same
for both simulations, but the differences are clearly seen by inspection of the
  RMS variability spectra. At 60\degr, the line is varying at three distinct
  wavelength ranges, but a more complex pattern is seen at 30\degr. Some of the
  additional variability will be caused by the observer seeing (at some phases)
  through the inner disc hole to the high-velocity gas close to a lower
  accretion hotspot. In both simulations, there are lines-of-sight that view a
  lower (beneath the disc plane) accretion stream through an upper accretion
  stream. The projected velocity differences between the two streams is
  generally greater when viewed from 30\degr\, inclination, so this is also
  expected to create additional variability.

\subsection{Offset magnetic axis}
\label{sec:A30}

\begin{figure*}
\begin{minipage}{180mm}
  \centering
  \begin{tabular}{c}
  \includegraphics[width=18cm,angle=270,clip=true]{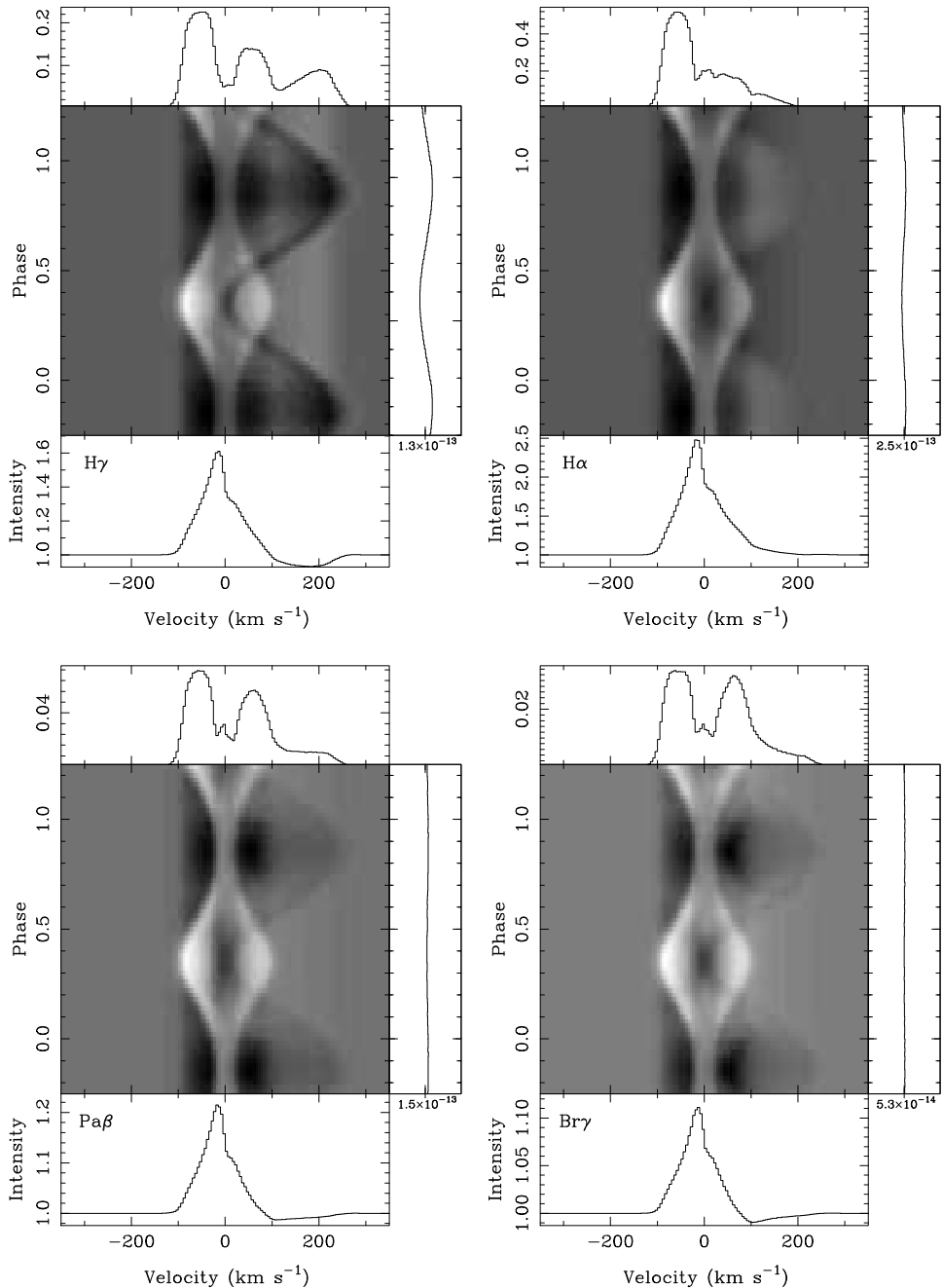}
  \end{tabular}
  
  \caption{Spectra from A30 model, viewed at inclination of $30\degr$. See
  Fig.~\ref{fig:d90} for a description. The variability shown in the greyscale
  has ranges (from top left) of 62, 112, 23 and 12 per cent of the mean
  spectrum in a given wavelength bin.}
  
  \protect\label{fig:a30}
\end{minipage}
\end{figure*}

Our A30 model is inspired by the MHD simulations of
\citet{2003ApJ...595.1009R}. The accretion streams are inclined by
10\degr\ with respect to the rotation axis, to represent a similarly
inclined dipolar magnetic field (see Section \ref{sec:example}). The accretion
columns subtend 30\degr\ in azimuth and, unlike the other models,
are not symmetric about their midplane. A stream exists above the disc
plane on one side of the star, and below the disc plane on the
opposite side (Fig.~\ref{fig:images1}d). There are therefore only two
hotspots on the stellar surface instead of the four that
are present in the other models.

The line profile variability through a rotation period reflects the
dramatic changes in the observer's view of the gas flow (Figs.~\ref{fig:a30},
~\ref{fig:snap_plots}b). The viewing angle of 30\degr\ 
(with respect to the disc normal and stellar rotation axes) keeps the
upper accretion hotspot visible at all phases, but each gas flow is
seen alternately towards, then away from the line of sight.
Additionally, the lower stream suffers from varying occultation by the
opaque accretion disc (affecting the low velocity component) and the
stellar disc (blocking the high velocity component close to the
obscured lower hotspot).

At phase 0.5 the upper stream is on the far side of the star and the
lower stream is at maximum visibilty through the disc's inner hole (Fig.~\ref{fig:images2}c).
All the hydrogen lines are composed of two distinct emission features
-- one redward of the rest wavelength and a larger one blueward.
Material close to the disc with extremely low velocities is only seen
in the upper stream and is visible throughout the rotation period so
the line centre is not particularly variable. The varying presentation
of the high-velocity gas in the upper stream dominates the quotient
spectra of Fig.~\ref{fig:a30}. Phase 0 sees the upper stream
towards the observer, with its maximum projection onto the stellar
disc. The blue continuum from the accretion hotspot is seen through
the highest velocity gas and so the deepest wavelength bin in the
redshifted absorption feature occurs at $+210$\kmsec\ in the \hg\ 
line. At \bg, the unheated stellar surface contributes most of the
continuum and the large volume of intermediate velocity gas seen over
the star creates an absoption minimum at only $+60$\kmsec. \ha\ is not
seen in absorption at any phase. 
There is a general trend for an enhancement of the line at two
wavelengths, symmetrically about the rest wavelength, which moves
alternately to lower and higher velocities through the rotation
period.

\begin{figure*}
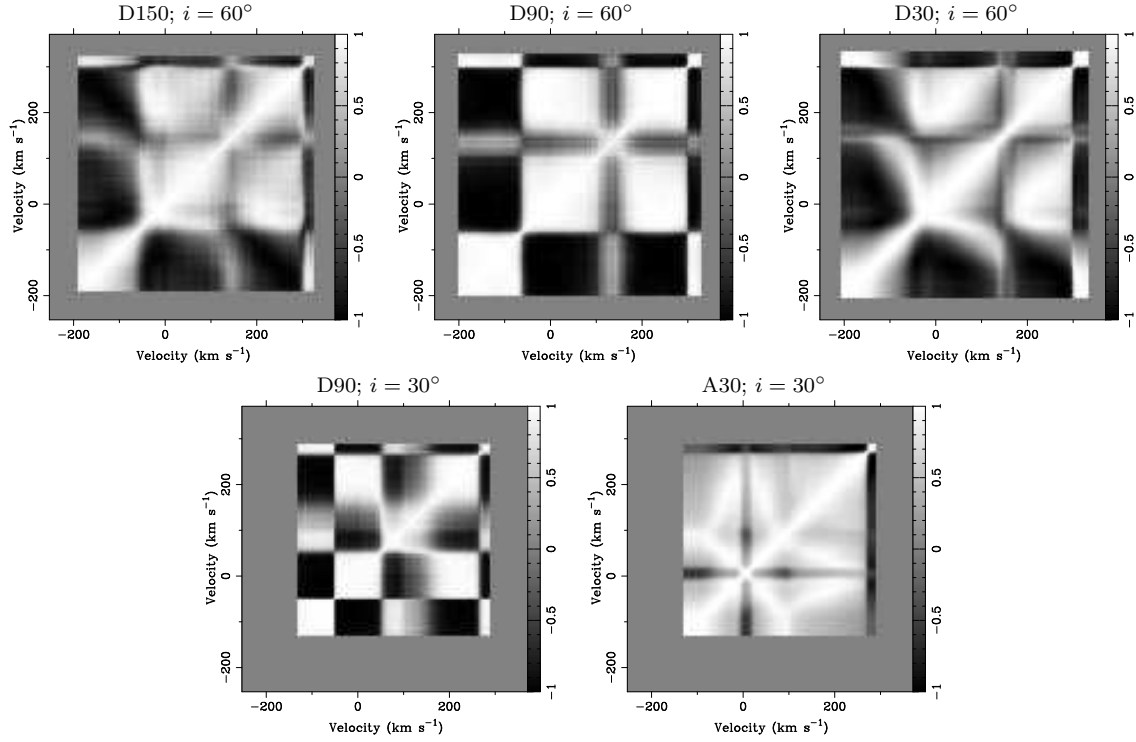

\begin{minipage}{180mm}
  \centering
\begin{tabular}{ccc}
  D150; $i=60\degr$   &      D90; $i=60\degr$ &  D30; $i=60\degr$ \\

  \includegraphics[width=4.7cm,angle=0,clip=true]{fig10_1a.eps} &
  \includegraphics[width=4.7cm,angle=0,clip=true]{fig10_1b.eps} &
  \includegraphics[width=4.7cm,angle=0,clip=true]{fig10_1c.eps} \\

\end{tabular}
\begin{tabular}{cc}
   D90; $i=30\degr$ &   A30; $i=30\degr$ \\

  \includegraphics[width=4.7cm,angle=0,clip=true]{fig10_2a.eps} &
  \includegraphics[width=4.7cm,angle=0,clip=true]{fig10_2b.eps} \\
  
\end{tabular}

  \caption{Images representing the auto-correlation function for the \hg\ line.
  The accretion model and the viewing inclination are shown above each image.
  The brightest shades represent the strongest correlation; the darkest shades
  are the strongest anti-correlations.}
  
  \protect\label{fig:acorr_hgamma}
  
\end{minipage}
\end{figure*}

\begin{figure*}
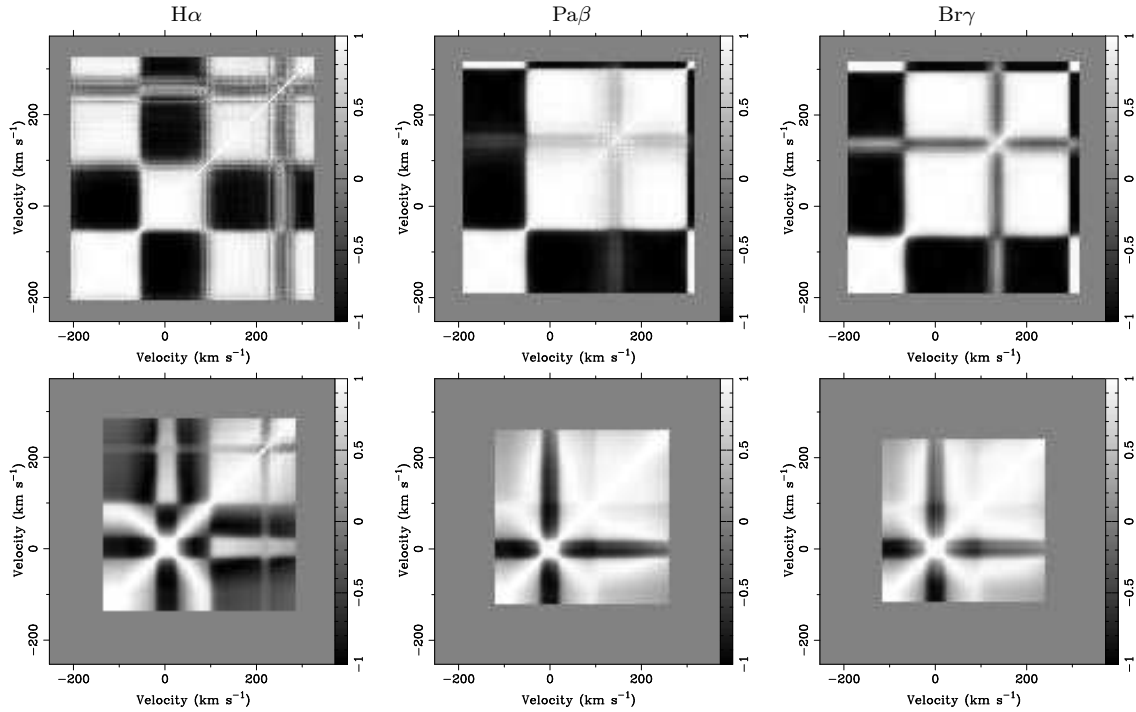

\begin{minipage}{180mm}
  \centering
  \begin{tabular}{ccc}
   \ha & \pb & \bg \\
  
  \includegraphics[width=4.7cm,angle=0,clip=true]{fig11_1a.eps} &
  \includegraphics[width=4.7cm,angle=0,clip=true]{fig11_1b.eps} &
  \includegraphics[width=4.7cm,angle=0,clip=true]{fig11_1c.eps}  \\
  
  \includegraphics[width=4.7cm,angle=0,clip=true]{fig11_2a.eps} &
  \includegraphics[width=4.7cm,angle=0,clip=true]{fig11_2b.eps} &
  \includegraphics[width=4.7cm,angle=0,clip=true]{fig11_2c.eps}  \\
  
  \end{tabular} \caption{Images representing the auto-correlation function for
  some spectral lines. Top row: D90 model at 60\degr\,inclination; bottom row:
  A30 at 30\degr\,inclination. Within each row, the columns are, from left to
  right, \ha, \pb\ and \bg. The brightest shades represent the strongest
  correlation; the darkest shades are the strongest anti-correlations.}
  \protect\label{fig:acorr_other}

\end{minipage}
\end{figure*}

\subsection{Correlation of variability}
\label{sec:correlation}

Observational studies of line profile variability in CTTS frequently use
cross-correlation images to interpret the often complex phenomena that occur
across profiles and between profiles from different transitions of species; a
technique pioneered by \cite{1994PhDT........21J} and
\cite{1995AJ....109.2800J}. We have computed auto-correlation (AC) images (see
Figs.~\ref{fig:acorr_hgamma} and \ref{fig:acorr_other}) for each line of our
model time-series, following the method of \citet{2000A&A...362..615O}. The
grey border of each image corresponds to the continuum, with white regions
describing positive correlation and black features indicating negative
correlation.

The \hg\ AC image for the D90 model at $i=60\degr$ shows a region of positive
correlation around the rest velocity of the line and at high redwards
($200-250$\kmsec) and most bluewards ($< 50$\kmsec) velocities.  There is a
`cross' of low correlation centred on $(+130,+130\kmsec)$ and strong
anticorrelation between the red side of the profile ($0-250$\kmsec) and the
blue side of the line ($< -70$ \kmsec). This anticorrelation is intrinsically
linked to the diametric nature of the accretion curtains: when one curtain is
in front of the star (leading to enhanced absorption), the other is behind the
star (leading to enhanced emission at blue velocities).  The `cross'
is related to the D90 RMS spectra in Fig.~\ref{fig:allExtents},
which shows a minimum at $\sim +130$ \kmsec. The discontinuity between
positive and negative correlation at $\sim -70$ \kmsec\ corresponds to
another minimum in the RMS spectra, and it can clearly be seen in the
grayscales that the intensity varies in antiphase on either side of this
velocity.

The D150 and D30 models at 60\degr\, inclination have similar features, and so
do the D90 models in the near infrared \pb\ and \bg\ lines. The D90 \ha\ line
  has an anti-correlation region between the rest velocity and velocities
  greater than $+100$\kmsec; this additional feature arises because \ha\ is the
  only line in the D90 $i=60\degr$ model that does not show an IPC profile. The
  same effect is seen in the \ha\ line of the A30 model, whose other lines are
  positively auto-correlated over most of their wavelength ranges.

\begin{figure*}
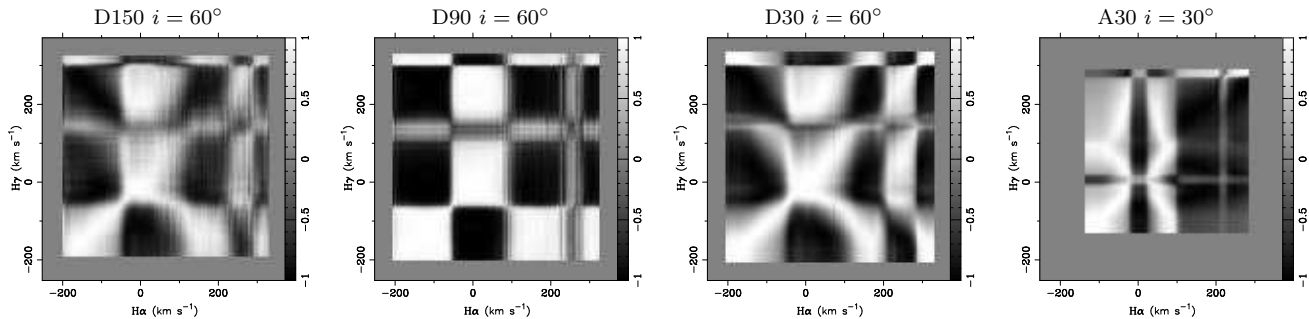

\begin{minipage}{180mm}
  \centering
  \begin{tabular}{cccc}
  D150 $i=60\degr$ &  D90 $i=60\degr$ & D30 $i=60\degr$ & A30 $i=30\degr$ \\
  \includegraphics[width=4cm,angle=0,clip=true]{fig12_a.eps} &
  \includegraphics[width=4cm,angle=0,clip=true]{fig12_b.eps} &
  \includegraphics[width=4cm,angle=0,clip=true]{fig12_c.eps} &
  \includegraphics[width=4cm,angle=0,clip=true]{fig12_d.eps} \\

  \end{tabular}
  
  \caption{Cross-correlation maps for the \ha\ and \hg\ spectral lines in the
  four accretion models. The model and the viewing inclination are shown above
  each image. The brightest shades represent the strongest correlation; the
  darkest shades are the strongest anti-correlations.}
  
  \protect\label{fig:xcorr}
\end{minipage}
\end{figure*}  

We have also computed cross-correlation (CC) images (with zero lag) of the \ha\
profiles with the \hg\ profiles for each of our models (Fig.~\ref{fig:xcorr}).
For the D90 model the cross-correlation image demonstrates that the
variability in the red sides of the \ha\ and \hg\ profiles is anticorrelated.
This anticorrelation is directly visible in the grayscale images of
Fig.~\ref{fig:d90}, with bright features (excess emission) on the red side of
\ha\ coinciding with increased absorption in \hg. A similar pattern is seen for
the D150 and D30 CC images. The A30 CC image shows that the red wing of the \ha\
profile is anticorrelated with practically the whole \hg\ line. This is a
consequence of the red wing varying for only fraction of the rotation period --
the same range of phases at which the upper hotspot is visible and suppressing the
emission line.

\section{Discussion}
\label{sec:discussion}

The illustrative models presented in this paper represent a first step
in quantifying the emission line profile variability observed in CTTS.
We did not adjust the simulation parameters to achieve agreement with
any existing data, and our choice of simple configurations has
produced line profiles with high levels of variation. It is therefore
most useful to consider the general characteristics seen in our
time-series spectra.

Our synthetic spectra typically predict levels of line profile variability that
are significantly greater than those observed.  Only the D150 model, which has
wide accretion curtains, shows variability that is comparable to published data
e.g.  \citet{2001A&A...378.1003S}.  The geometry with the least symmetry (A30)
leads to line profiles that change substantially in shape and intensity from
one rotation phase to another. This geometry was selected to represent the
structures indicated by recent MHD calculations \citep{2003ApJ...595.1009R},
but it appears that the line profiles predicted by such a magnetosphere are
sufficiently at odds with spectroscopic observations that this form of MHD
model may be rejected.

MHD simulations must balance the interactions of the accretion disk
with a magnetic field that includes large scale structure (e.g. a
dipolar component) with local irregularities. Both accretion flows
onto the star and magnetically launched outflows
\citep[e.g][]{1997ApJ...489..890M} must be derived self-consistently
to explain observed spectral features. CTTSs are probably more fully
enveloped in accreting gas than our A30 model proposes, but equally,
the broad symmetric curtains of our D150 and D90 models are too
regular to represent the circumstellar environment. Better agreement
might have been reached by keeping the high density streams, but
elsewhere setting a lower mass accretion rate that would increase the
azimuthal filling factor of the magnetosphere.

The RMS spectra of our simulations often show several distinct minima,
corresponding to velocities where the emission from line-of-sight
iso-velocity surfaces remains constant. This pattern of variability
is not often seen in the observations, although \ha, \hb, and Na\,{\sc
  i} doublet variance profiles of SU~Aur \citep{2000A&A...362..615O}
do show multiple regions of mininum variability. More typically the
line variability is most pronounced either on the red
\citep{2001A&A...378.1003S} or blue \citep{2001AJ....122.3335A} sides
of the profile.  Occassionally a bimodal distribution of variability
is seen across the emission lines, with equal variability in the red-
and blue-line-wings and a minimum at the line centre
\citep{2002ApJ...571..378A}. Given that our models neglect any outflow
component, our RMS spectra typically indicate strongest variability on
the red side of the model profiles. However the A30 model, with its
narrow curtains, shows that blue-side line variability can be dominant
(particularly at \ha) even with an infall geometry.

Cross- and auto-correlation images are widely used as a diagnostic of
line profile variability in CTTS (e.g. \citealt{1995AJ....109.2800J,
  2000A&A...362..615O, 2001AJ....122.3335A, 2002ApJ...571..378A}). A
square, cross-shaped auto-correlation function is often seen in the
lower-order Balmer lines (e.g.  \citealt{1995AJ....109.2800J,
  2002ApJ...571..378A, 2000A&A...362..615O}) and this has been
interpreted as resulting from viewing the the magnetosphere at low
inclinations \citep{2002ApJ...571..378A}. We find such a pattern in
our AC images for essentially all our models (see
Figs.~\ref{fig:acorr_hgamma},~\ref{fig:acorr_other}), indicating that the AC images may be of limited
diagnostic potential. The cross-correlation images at zero lag (see
Fig.~\ref{fig:xcorr}) show a variety of structures, but we find no
obvious similarities with the published CC images.

Interestingly, while the variability of the emission line profiles in
our simulations is excessive, the ranges of continuum flux
(Table~\ref{table:models}) are comparable with, or even lower than,
results from photometric surveys, where $V$ band magnitudes are seen
to change periodically by up to $\sim1$~mag
\citep[e.g.][]{1993A&A...272..176B}. In the models
presented here, any smoothing of the azimuthal magnetospheric
structure (broadening of the curtains for example) in order to reduce
line profile variablity, will lead to an equal smoothing of the
surface hot spots (and a corresponding reduction in continuum
variability). The simultaneous satisfaction of the dual spectroscopic
and photometric observational constraints represents a challenging
problem.

\section{Conclusions and future work}
\label{sec:conclusions}

We have presented the first 3D radiative transfer calculations of
non-axisymmetric accretion onto classical T~Tauri stars. Hydrogen line
profiles throughout the stellar rotation period are shown for three
systems that have accretion confined to curtains filling part of a
dipolar magnetosphere. Three of the models had curtains that were
symmetric about the disc plane, and the fourth accreted from above the
disc on one side of the star, and from below on the other side. The
latter configuration was suggested by MHD simulations of stars with an
inclined dipolar field, but we find that our model creates line
profiles that are excessively variable when compared to observations.
We suggest that CTTS magnetospheres may have high filling factors of
accreting gas, but with local enhancements that may be in the form of
streams confined in azimuth.

The simple models presented here are, of course, unable to reproduce
the wide range of variable phenomenon observed. Nonetheless, some
gross characteristics of the line profile variability are reflected in
the models, including the form of the variability's velocity
distribution and the auto-correlation images. In the future we will
investigate the line and continuum diagnostics of non-static accretion
models.  Time-series spectroscopy of AA~Tau by
\citet{2003A&A...409..169B} already hints at changes in the global
accretion rate on a time-scale of hours. Using a simple freefall model
we should be able to examine how a change in the mass-accretion rate
affects the line profiles and continuum flux as it propagates down the
field lines. Our self-consistent model for the reprocessed accretion
power should enable us to determine the lag between the line and
continuum variability.

Furthermore, the observation of anticorrelation betwen the blue and
red side variability of \ha\ in SU~Aur has been interpreted using a
dipole offset model with simultaneous dipolar accretion and an outflow
\citep{1995ApJ...449..341J}. The inclusion of a wind component in our
geometry will enable us to investigate such scenarios and make
quantative predictions regarding the line variability.

Finally, the \textsc{torus} code includes computation of the Stokes
vectors for the simulated photon packets, and can measure linear
polarization by scattering events. Addition of a magnetic field
configuraton to our simulation grid would allow circular
spectropolarimetry to be simulated.  This technique has already been
used observationally to show CTTS emission lines originate in structures with a net
magnetic field \citep{1999ApJ...510L..41J,TEMPsym04b}, so self-consistent
modelling of line intensity and polarization would allow more
stringent tests of magnetospheric accretion models.

\section*{Acknowledgements}

RK is funded by PPARC standard grant PPA/G/S/2001/00081. 

\bibliography{nhs} \bibliographystyle{mn2e}

\end{document}